\documentclass[pre,eqsecnum,showpacs,amssymb]{revtex4}
\usepackage{amsmath,graphicx}
\usepackage{epsfig}
\usepackage{bm}

\begin{document}

\title{Driven  translocation of a polymer:  fluctuations at work}

\author{J. L. A. Dubbeldam$^{2}$, V. G. Rostiashvili$^1$, A.
Milchev$^{1,3}$, and T.A. Vilgis$^1$}
\affiliation{$^1$ Max Planck Institute for Polymer Research, 10 Ackermannweg,
55128 Mainz, Germany\\
$^2$ Delft Institute of Applied Mathematics, Delft University of Technology
2628CD Delft, The Netherlands\\
$^3$ Institute for Physical Chemistry, Bulgarian Academy of Sciences, 1113
Sofia, Bulgaria}

\begin{abstract}
The impact of thermal fluctuations on the translocation dynamics of a polymer
chain driven through a narrow pore has been investigated theoretically and by
means of extensive Molecular-Dynamics (MD) simulation. The theoretical
consideration is based on the so-called velocity Langevin (V-Langevin) equation
which determines the progress of the translocation in terms of the number of
polymer segments, $s(t)$, that have passed through the pore at time $t$ due to
a driving force $f$. The formalism is based only on the assumption that, due to
thermal fluctuations, the translocation velocity $v=\dot{s}(t)$ is a Gaussian
random process  as suggested by our MD data. With this in mind we have derived
the corresponding Fokker-Planck equation (FPE) which has  a {\em nonlinear}
drift term and diffusion term with a {\em time-dependent} diffusion coefficient
$D(t)$. Our
MD simulation reveals that the driven translocation process follows a
{\em super}diffusive law with a running diffusion coefficient $D(t) \propto
t^{\gamma}$ where $\gamma < 1$. This finding is then used in the numerical
solution of the FPE  which yields an important result: for comparatively small
driving forces fluctuations facilitate the translocation dynamics. As a
consequence, the exponent $\alpha$ which describes the scaling of the mean
translocation time $\langle \tau \rangle$ with the length $N$ of the polymer,
$\langle \tau \rangle \propto N^{\alpha}$ is found to diminish. Thus, taking
thermal fluctuations into account, one can explain the systematic discrepancy
between theoretically predicted duration of a driven translocation process,
considered usually as a deterministic event, and measurements in computer
simulations.

In the non-driven case, $f=0$, the translocation is slightly subdiffusive and
can be treated within the framework of fractional Brownian motion (fBm).
\end{abstract}


\maketitle

\section{Introduction}
The polymer translocation through narrow pores is one of the most striking
single-molecule experiments motivated mainly by the DNA transport across cell
membranes and the possibility of DNA sequencing\cite{Meller,Laan}. In nearly two
decades of intensive investigations considerable headway has been achieved by a
variety of experimental, theoretical and simulational studies\cite{Milchev}. 
In most cases of experimental and simulation studies one deals with a polymer
chain that is driven by external electrical field which acts on the monomers
passing through the nanopore from the {\em cis}- to the {\em trans}-side of the
separating membrane \cite{Meller}. Some of the very basic questions one tries
to answer concern the time $\tau$ it takes the chain to thread and translocate
between the two half-spaces of the setup and in particular, how $\tau$ depends
on chain length $N$ and applied force $f$.

The simplest theoretical approach has considered this process as one-dimensional
biased diffusion (in terms of the translocated number of segments $s$) over an
entropic barrier \cite{Sung,Muthu}. Such an approach obviously suffers from some
inconsistency since the mean time of the driven translocation $\langle \tau
\rangle$ turns to be smaller than the relaxation (Rouse) time of the chain
$\tau_{R}$ so that the chain has not enough time to equilibrate and experience
the global entropic barrier. Later, it was realized that the chain responds to
pulling at first locally in the vicinity of the nanopore with a tensile force
spreading away from the separating membrane. This scenario has been treated
within the linear response theory (that is, for relatively weak driving forces)
and the corresponding memory function has been derived explicitly
\cite{Panja_1,Panja_2}. For arbitrary strong driving forces, an interesting
approach based on the notion of tensile force propagation along the chain
backbone has been suggested by Sakaue \cite{Sakaue_1, Sakaue_2,Sakaue_3}.
Sakaue's idea has been used and worked out very recently in other theoretical
treatments \cite{Dubbeldam_1,Ikonen}.

Despite increased insight in the nature of the translocation process, however,
current understanding of the translocation dynamics is still far from
satisfactory. So, among other things, there exists a non-negligible systematic
discrepancy between the theoretical (analytical) predictions concerning the main
scaling exponents that characterize translocation dynamics and the data provided
by computer experiments, mainly Molecular Dynamics (MD) simulation studies
\cite{Milchev}.

Recently \cite{Dubbeldam_1} we suggested a theoretical description based on
the tensile (Pincus) blob picture of a pulled chain and the notion of a tensile
force propagation, introduced by Sakaue \cite{Sakaue_1,Sakaue_2,Sakaue_3}.
Assuming that the local driving force is matched by a drag force of equal
magnitude, (i.e., in a quasi-static approximation), we derived an equation of
motion for the tensile front position. This enables one to calculate the
deterministic dependence (i.e., without taking into account fluctuations) of the
translocation coordinate as a function of time $M(t)$. One can obtain the
scaling law for the mean translocation time $\langle \tau \rangle$ vs. chain
length $N$, i.e., $\langle \tau \rangle \propto N^{\alpha}$, where $\alpha$ is
the translocation exponent. The total translocation process consists of two
stages. First, the tensile force propagates so that more and more polymer
segments get involved in the moving domain. Depending on the driving force $f$,
which is applied to the segment in the pore, the moving domain attains different
shapes: ``trumpet'', ``stem-trumpet'', or a ``stem'',  for weak, intermediate
and strong forces, respectively. Once the last polymer segment on the {\it
cis}-side of the pore gets involved, the velocity of the moving domain
approaches a stationary value. After that the second stationary stage sets in
and the rest of the chain is sucked into the pore with constant velocity.

We have shown \cite{Dubbeldam_1} that for the ``trumpet'' case, the
characteristic duration (in the case of Rouse dynamics) of the first and second
stages goes as $\tau_1 \propto N^{1+\nu}/f$ and $\tau_2 \propto
N^{2\nu}/f^{1/\nu}$ respectively (where $\nu \approx 3/5$ is the Flory
exponent), so that at ${\tilde f}_R \stackrel{\rm def}{=} a N^{\nu} f/k_B T \gg
1$ the first stage time dominates. As a consequence, with growing chain length
$N$ and force $f$, the translocation exponent $\alpha$ also increases from
$\alpha = 2\nu \approx 1.19$ to $\alpha = 1 + \nu \approx 1.59$. These
predictions are supported by our simulation findings \cite{Dubbeldam_1} as well
as by results of Lehtola {\it et al.} \cite{Lehtola_1, Lehtola_2,Lehtola_3} but
differ from the results of Luo {\it et al. } \cite{Luo}. The tensile force
propagation phenomenon treated within the iso-flux trumpet model also leads to
$\alpha = 1 + \nu$ \cite{Grosberg}.

Unfortunately, the MD-simulation results yield systematically smaller values for
the translocation exponent $\alpha$. For example, in our MD-simulations $\alpha
\approx 1.33$ for strong forces and $\alpha \approx 1.06$  for weak forces
(distinct from the theoretical predictions $\alpha = 1 + \nu \approx 1.59$, and
$\alpha = 2\nu \approx 1.18 $ respectively). It is important to determine the
origin of this inconsistency since apparently there is something missing in the
afore mentioned theoretical consideration which gives room for speculations. For
example, in a paper by Ikonen et at. \cite{Ikonen}, the model based on the idea
of tensile force propagation \cite{Sakaue_1, Sakaue_2,Sakaue_3} and the role of
pore-polymer friction has been numerically investigated. The authors argue
that the
theoretical value for the exponent, $\alpha = 1 + \nu$, may be seen only for
very long chains whereas for the chain lengths used in real experiments or
simulations the effective exponent $\alpha$ could be approximately 20\% smaller.

In this work we consider theoretically and by means of MD-simulations how
fluctuations of the translocation coordinate (such fluctuations have been
ignored in the quasi-static approximation as in the other deterministic
treatments) affect the forced translocation dynamics. Indeed, the role of
thermal fluctuations is by no means self-evident. A recent publication
\cite{Sakaue_4} argues that only fluctuations related to the initial
distribution of segments are essential whereas thermal fluctuations have a minor
effect. Moreover, the authors  argue \cite{Sakaue_4} that the translocation
exponents are not affected by the fluctuations in the initial conditions.

In the
present investigations we demonstrate that thermal fluctuations may
facilitate the translocation dynamics so that the effective translocation
exponent $\alpha$ becomes smaller. To this end in Section \ref{Theory} we derive
a Fokker-Planck equation (FPE) for the translocation coordinate $s$ probability
distribution function
$W(s, t)$. This FPE, which contains a nonlinear drift term and a time-dependent
diffusion coefficient $D(t)$, is then solved numerically demonstrating a
significant fluctuation-induced facilitation of the translocation
dynamics. In Section \ref{Sim} we then present our simulation results. First, we
prove that the translocation velocity under differently strong driving forces is
a random process with a Gaussian distribution which is the only conjecture we
need for our theory. Then we calculate the velocity autocorrelation function
(VAF) which reveals an oscillatory behavior and a long-time tail. This gives
rise to a time-dependent diffusion coefficient $D(t) \propto t^{\gamma}$, where
the exponent $\gamma < 1$, indicating a {\em super}-diffusively driven
translocation. We end this report with a brief summary of conclusions in
Section \ref{Summary}. Some technical details relegated to the Appendices
\ref{App_1}, \ref{App_2}.

\section{From  Langevin to Fokker-Planck equation}
\label{Theory}

In order to allow for fluctuations of the translocation coordinate one should
consider the problem within the framework of the corresponding Fokker-Planck
(FPE) equation which governs the probability distribution function (PDF) $W(s,
t)$ that exactly $s$ segments have passed the pore at time $t$. By so doing we
follow our earlier approach ref. \cite{Dubbeldam_2} where the case of undriven
translocation was treated. We focus  here on the general case when the drift
term in FP-equation is essential

Our approach is based on the assumption that the translocation coordinate
$s(t)$, which measures the number of segments heaving reached the {\em
trans}-side at time $t$, is a random process  governed by the so-called
V-Langevin (where V stands for ``velocity'') equation
\begin{eqnarray}
 \dfrac{d s(t)}{d t} = v \left(s(t) \right)
\label{Langevin}
\end{eqnarray}
This equation has been discussed extensively by Balescu \cite{Balescu} mainly in
the context of plasma dynamics. The translocation velocity $v \left(s(t)
\right)$  in Eq. (\ref{Langevin}), which depends on the trajectory $s(t)$, is
assumed to be a Gaussian process  with given first moment
\begin{eqnarray}
 \left\langle v (s (t)) \right\rangle \stackrel{\rm def}{=} \left.
K(s)\right|_{s = s(t)} \stackrel{\rm def}{=} - \dfrac{1}{\xi_0} \left.
U^{\prime}(s)\right|_{s = s(t)}
\label{First}
\end{eqnarray}
and a two-point correlation function
 \begin{eqnarray}
   \left\langle \left[ v(s (t_1)) - \langle v( s (t_1)) \rangle \right] \left[
v( s (t_2))
- \langle  v(s (t_2)) \rangle \right]\right\rangle  \stackrel{\rm def}{=} G
(t_1,
t_2)
\label{Second}
 \end{eqnarray}.

In Eq. (\ref{First}) the function $U(s)$ stands for the effective potential in
the translocation coordinate space $s$ while $\xi_0$ is the corresponding
friction coefficient. The form of the averaged velocity $K(s)$ as well as the
effective potential $U(s)$ we will find below using the {\it correspondence
principle}. This principle states that in the limit of very small fluctuations
the FPE solution reproduces the deterministic (or quasi-static) solution
$\left\langle s(t) \right\rangle = M (t)$ which was derived in ref.
\cite{Dubbeldam_1}. The case of colored Gaussian statistics as well as the
behavior of the VAF  $G (t_1,t_2)$ will be studied by MD-simulation in Sec.
\ref{Sim}. This will be considered as the required dynamic  input  while solving
the corresponding FPE. 

Based on the V-Langevin Eq. (\ref{Langevin}), one could derive the
corresponding FPE for the PDF  $W(s, t)$. The latter is defined as follows
\begin{eqnarray}
 W(s, t) = \langle \delta (s - s(t))\rangle 
\label{PDF}
\end{eqnarray}
where $\langle \dots \rangle$ means an average over the Gaussian velocity field
fluctuations with a first moment and two-point correlation function given by Eq.
(\ref{First}) and Eq. (\ref{Second}), respectively. The detailed derivation is
given in Appendix \ref{App_1}. The resulting FPE reads
\begin{eqnarray}
 \dfrac{\partial W (s, t)}{\partial t}  = - \dfrac{\partial}{\partial s}
[K (s) W (s, t)] + {D (t)} \: \dfrac{\partial^2}{\partial s^2}  \: W (s, t)
\label{FP_Final}
\end{eqnarray}
where the time dependent diffusion coefficient
\begin{eqnarray}
D (t) = \int_{0}^{t} G(t, \tau)
d \tau.
\label{D}
\end{eqnarray}
It should be stressed that Eq. (\ref{FP_Final}) is rather general. Indeed, no
assumptions regarding, for example, the validity of the fluctuation-dissipation
theorem are made in the course of derivation in contrast to the case of the
generalized Langevin equation approach \cite{Cherail,Panja}. Moreover, the drift
and diffusion terms are totally independent (i.e. the Stokes-Einstein relation
does not necessarily hold). In other words, this FPE is valid even for a highly
non-equilibrium driven translocation processes.

\subsection{What is the form of $K (s)$?}

As mentioned above, the form of $K (s)$ can be fixed by making use of the
correspondence principle which states that in the limit of zero fluctuations the
FPE reproduces the deterministic solution for the first moment, $\left\langle
s(t)\right\rangle = M (t)$, derived in ref. \cite{Dubbeldam_1}, i.e.
\begin{eqnarray}
 M(t) = \left( c_0 {\widetilde f}_a {\widetilde t}\right)^{\beta}
\label{M}
\end{eqnarray}
where the dimensionless force ${\widetilde f}_a = a f/T$ and time ${\widetilde
t}= t/\tau_0$ with $a$, $T$ and $\tau_0 = a^2 \xi_0/T$ standing for the Kuhn
segment length, temperature, and the microscopic characteristic time,
respectively. In Eq. (\ref{M}) $c_0$ is a constant of the order of unity and the
exponent $\beta$ in the limit of strong driving forces reads $\beta = 1/(1 +
\nu)$ \cite{Dubbeldam_1}.

\begin{figure}[ht]
 \includegraphics[scale=0.5,angle=270]{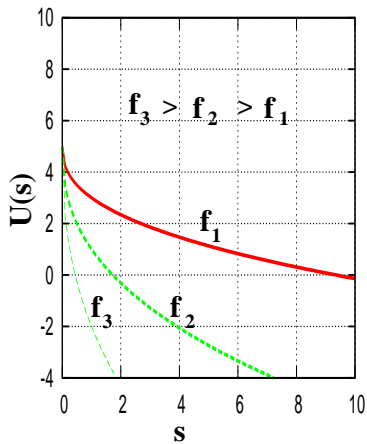}
\caption{Effective potential $U (s)$ ( given by  Eq.( \ref{U_Potential})) as a
function of the translocation coordinate $s$ at different forces: $f_3 > f_2 >
f_1$.}
\label{Landscape}
\end{figure}

In this limit $G(t, \tau) = 0$, $D (t) = 0$ and Eq. (\ref{FP_Final}) reduces to
\begin{eqnarray}
 \dfrac{\partial W (s, t)}{\partial t}  = - \dfrac{\partial}{\partial s} [K (s)
W(s, t)]
\label{Zero_Fluctuation_1}
\end{eqnarray}
On the other hand, in this case $ W (s, t) = \delta (s - M(t))$ and Eq.
(\ref{Zero_Fluctuation_1}) is equivalent to
\begin{eqnarray}
 \dfrac{\partial}{\partial t} \delta (s - M(t)) &=& - K (M(t)) \;
\dfrac{\partial}{\partial s} \: \delta (s - M(t)) \nonumber\\
&=& - {\dot M} \: \dfrac{\partial}{\partial s} \: \delta (s - M(t))
\end{eqnarray}
where in the first line  we have used that $K (s) \delta (s - M(t)) = K (M(t)) 
\delta (s - M(t))$, and in the second line the chain rule has been used.
As a result the FPE is equivalent to the deterministic equation
\begin{eqnarray}
 \dfrac{d}{d t} M(t) = K(M(t))
\label{Deterministic}
\end{eqnarray}
It is easy to show that in order to restore the solution, Eq. (\ref{M}), the
function $K (s)$ in Eq. (\ref{Deterministic}) should be of the form
\begin{eqnarray}
 K (s) = \dfrac{\beta c_0 f}{a \xi_0} \: s^{-(1 - \beta)/\beta}
\label{K}
\end{eqnarray}
 It is also instructive to reconstruct the form of the effective potential
$U(s)$, given by Eq. (\ref{First}). Namely, from Eq. (\ref{First}) and Eq.
(\ref{K}) one has $U^{\prime}(s) = - (\beta c_0 f/a )  \: s^{-(1 -
\beta)/\beta}$. As a result, the effective potential that governs the
translocation reads
\begin{eqnarray}
 U (s) = U_0 - \dfrac{\beta c_0 f}{a (2 - 1/\beta)} \: s^{2 - 1/\beta}
\label{U_Potential}
\end{eqnarray}
i.e., it has the form of the slanting plane shown in Fig. \ref{Landscape}.

\subsection{Two ways  to calculate $D(t)$ by using MD-data}

The behavior of the diffusion coefficient $D(t)$ which is necessary as an input
for the solution of the FPE, could be found using a MD-simulation by two
methods. First, one could directly calculate the VAF $G(t, \tau)$ and then
integrate it over the time according to Eq. (\ref{D}). On the other hand, there
is a simple relationship between the time dependent variance $\left\langle
\Delta s^2 (t) \right\rangle \stackrel{\rm def}{=} \langle [ s(t) - \langle s(t)
\rangle]^2\rangle$ and $D (t)$. Indeed, first of all $ s(t) = s_0 + \int_{0}^{t}
v (s(t_1)) d t_1$ , i.e. $\langle s(t) \rangle = s_0 + \int_{0}^{t} \langle v
(s(t_2))\rangle d t_2$. Thus,
\begin{eqnarray}
 \left\langle  [s (t) - \langle s(t) \rangle]^2 \right\rangle &=&  \int_{0}^{t}
d t_1 \int_{0}^{t} d t_2 \left\langle \delta v (s(t_1)) \delta v
(s(t_2))\right\rangle\nonumber\\
&=& \int_{0}^{t}d t_1 \int_{0}^{t} d t_2 \: G (t_1, t_2)
\label{Variance}
\end{eqnarray}
where, as before, $\delta v (s(t)) = v (s(t)) - \langle v (s(t)) \rangle$.
Finally, differentiation of $\left\langle \Delta s^2 (t)\right\rangle
\stackrel{\rm def}{=} \langle [ s(t) - \langle s(t)\rangle]^2\rangle$ in Eq.
(\ref{Variance}) leads to
\begin{eqnarray}
 \dfrac{d \left\langle \Delta s^2 (t)\right\rangle}{2 d t} = \int_{0}^{t} \: G
(t, \tau) d \tau = D(t)
\label{D_2}
\end{eqnarray}
where we have used that $G (t_1, t_2) = G (t_2, t_1)$. In Section \ref{Sim} we
will show that the mean squared displacement $\left\langle \Delta s^2(t)
\right\rangle$, calculated by the double integration of the  VAF  (see Eq.
(\ref{Variance})) over time, and by the direct simulation, gives closely
matching results.

\subsection{Numerical Solution of the FP-equation}
\label{Theory_Numerical}

The FPE, Eq. (\ref{FP_Final}), can be easily solved numerically assuming
reflection-adsorption boundary conditions and some reasonable guess
about the form of time-dependent diffusion coefficient $D(t)$. The direct
inspection of simulation data, given in Sec. \ref{Res}, shows that the VAF
$G(t_1, t_2)$ has a long-time tail, i.e., correlations persist in time. This,
according to Eq. (\ref{D_2}), leads to a time-dependent diffusion coefficient
$D(t)$ which could be approximated by a power law
\begin{eqnarray}
 D (t) = \dfrac{d_0}{\tau_0} \: {\tilde t}^{\gamma}
\label{Running_D}
\end{eqnarray}
where $d_0$ is a constant and the exponent $\gamma < 1$. Before we proceed
further with the results of numerical solution, let us give a short discussion
of the small-noise expansion of the FPE, Eq. (\ref{FP_Final}), where the
diffusion coefficient is a small time-dependent function, i.e., $D(t) =
(\varepsilon^2/2) D_0 (t)$, with $D_0 (t) \propto {\tilde t}^{\gamma}$.
Following the book of Gardiner \cite{Gardiner}, in the Appendix~\ref{App_2} we
give a more extensive discussion of the small-noise expansion around the
deterministic solution. In particular, for the first moment this expansion
yields
\begin{eqnarray}
  \left\langle s (t) \right\rangle = \left(c_0
{\tilde f} {\tilde t}\right)^{\beta}  + \dfrac{\varepsilon^2 {\tilde
d_0}}{{\tilde f}^{\beta}} \: {\tilde
t}^{1 + \gamma - \beta}
\label{YY_Final}
\end{eqnarray}
As is evident from Eq. (\ref{YY_Final}), the fluctuations are
responsible for speeding of the translocation process. The corresponding
fluctuation increment is of the $\varepsilon^2$-order (the same as the diffusion
coefficient, $D(t) = \varepsilon^2 D_0(t)/2$, itself)  and goes with time as
$t^{1+\gamma - \beta}$. In order to estimate the exponent
$1+\gamma - \beta$, one needs the exponent $\gamma$. One could use the data for
$\left\langle \Delta s^2 (t)\right\rangle$, given by Bhattacharya et al.
\cite{Bhattacharya} where it was reported that $\left\langle \Delta s^2
(t)\right\rangle \sim t^{1.44}$. As a result, according to Eq. (\ref{D_2}), the
(reduced) diffusion coefficient $D_{0} (t) \sim t^{0.44}$, i.e. $\gamma = 0.44$.
This estimate leads (taking also into account that $\beta = 1/(1+\nu)
\approx 0.63$)  to the behavior of fluctuation increment $ \sim
t^{0.81}$ which facilitates the translocation dynamics as $ \left \langle s (t)
\right\rangle = (c_0
{\tilde f} {\tilde t})^{0.63} + (\varepsilon^2 {\tilde d_0}/{\tilde
f}^{0.63}) \:  {\tilde t}^{0.81} $. In the general case, an effective exponent
enhancement is possible, if at least $1 + \gamma - \beta > \beta$, i.e.,
\begin{eqnarray}
 \gamma > 2 \beta - 1
\label{Enhancement}
\end{eqnarray}
Thus, there is a lower limit for the $\gamma$ value which ensures the
effective exponent (in the $\langle s(t)\rangle$ vs. $t$ dependence)
enhancement. For example, if $\beta = 1/(1+\nu)$ then $\gamma >
(1-\nu)/(1+\nu) \approx 0.26$. 

The relationship, Eq. (\ref{Enhancement}), can be also readily obtained by the
following arguments. If $\langle \Delta s^2 \rangle \propto t^{\theta}$, then
the first two moments could be estimated as $\langle s \rangle \propto
t^{\theta/2}$ and $\langle s^2 \rangle \propto t^{\theta}$. But in the case of
quasi-static (no fluctuations) approximation $\langle s \rangle \propto
t^{\beta}$ and, provided that $\theta/2 > \beta$, some slope enhancement owing
to fluctuations takes place. Taking into  account that according to Eq.
(\ref{D_2} ) $\theta = \gamma + 1$, one immediately arrives at Eq.
(\ref{Enhancement}).

\begin{figure}
\vspace{0.7cm}
 \includegraphics[width=8.0cm]{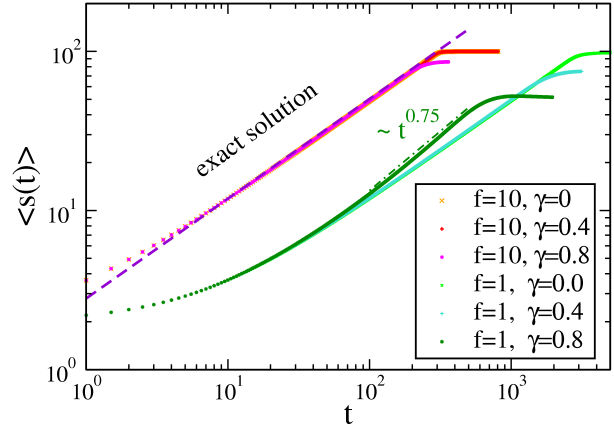}
\caption{The result of numerical solution of Eq. (\ref{FP_Final}) for two
different driving forces, $f=1$ (lower curves), $f=10$ (upper curves), and
exponents $\gamma=0, 0.4, 0.8$. The dashed line corresponds to the
quasi-static approximation with a slope $\beta = 1/(1+\nu) \approx 0.63$.
Taking into account fluctuations with $D(t) \propto t^{0.8}$ leads to a
slope (or, an effective exponent) enhancement up to $0.75$. The chain length $N
= 100$.}
\label{S_vs_time}
\end{figure}

Figure \ref{S_vs_time} demonstrates the  numerical solution of the FPE, Eq.
(\ref{FP_Final}), for the first statistical moment, $\langle s(t)\rangle $, as a
function of time (the chain length $N = 100$). It can be seen that for a
relatively strong force
\begin{figure}[htbp]
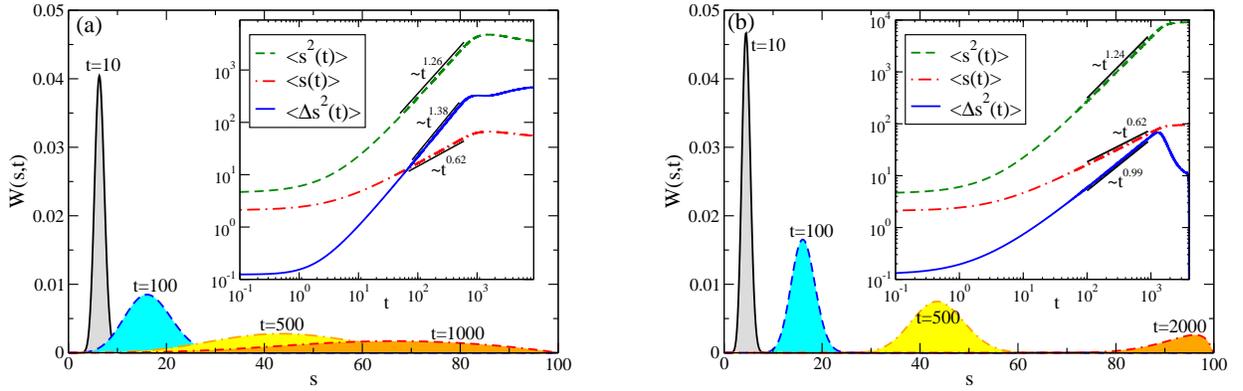

\includegraphics[width=7.5cm]{FIG3AFINAL.eps}
\hspace{1.0cm}
\includegraphics[width=7.5cm]{FIG3BFINAL.eps}
\caption{ (a) The PDF function $W(s, t)$ for $f = 1$, $\gamma = 0.4$ and
(inset) the corresponding statistical moments $\langle s(t)\rangle$, $\langle
s^2(t)\rangle$, $\langle \Delta s^2(t)\rangle$ . (b) The  same but for the
exponent $\gamma = 0$.}
\label{W_Distributions}
\end{figure}
(upper set of curves for $f=10$), the fluctuation has little or no effect on the
$\langle s(t)\rangle$ behavior. For weaker force, e.g., for $f = 1$, shown by
the lower set of curves, and relatively large exponent $\gamma = 0.8$, one can
see a clear slope enhancement, up to $0.75$. Thus, one may argue that under
relatively small driving forces fluctuations facilitate the translocation
dynamics.

\begin{figure}
\vspace{0.7cm}
 \includegraphics[width=8cm]{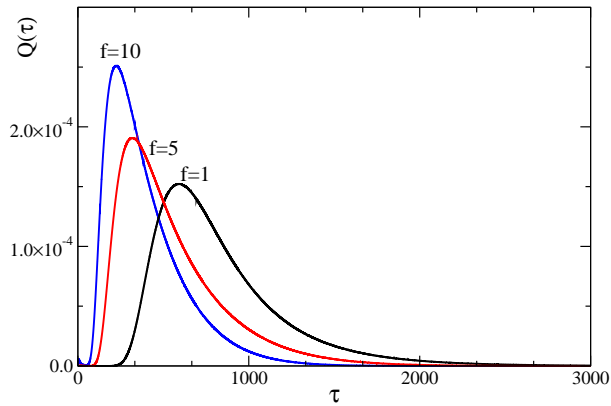}
\caption{The first-passage time distributions (FPTD) $Q(t)$  calculated for
$\gamma = 0.4$ and different forces $f = 1, 5, 10$.}
\label{Q}
\end{figure}

The results of numerical solution for the PDF $W(s, t)$ as well as the
corresponding statistical moments, $\langle s(t)\rangle$ , $\langle
s^2(t)\rangle$, $\langle \Delta s^2(t)\rangle$, are shown in Fig.
\ref{W_Distributions} for different parameters. As one might expect, the time
dependence of the variance follows the law $\langle \Delta s^2(t) \rangle
\propto t^{\theta}$, where the exponent $\theta = \gamma + 1$ in accordance with
Eq. (\ref{D_2}). In other words, for the running diffusion coefficient, $D(t)
\propto t^{\gamma}$,   with $\gamma < 1$, the exponent $1 < \theta < 2$ and one
finds a case of {\it superdiffusion}. It is of interest that, due to the
external force (in this case $f = 1$) and the adsorption boundary condition
($W(s = N,t) = 0$), the PDF $W(s, t)$ becomes narrower at a later stage of the
translocation. This results in a non-monotonic behavior of the variance $\langle
\Delta s^2(t)\rangle$ for $\gamma =0.4$, and especially for $\gamma =0$ (see
Fig. \ref{W_Distributions}). These numerical findings for the statistical
moments are qualitatively consistent with our MD-results given in Fig.
\ref{fig:figureS2b}. Eventually, the {\it first passage time probability
distribution} (FPTD), $Q(t) = - (d/d t) \int_{0}^{N} \: W(s, t) ds $,  for
different forces ($f = 1, 5, 10$) is depicted in Fig.~\ref{Q}.

\subsection{Continuous Time Random Walks (CTRW) approach and beyond}

It is pertinent to note that the interpretation of the {\it superdiffusion} in
terms of the time persistent correlations in the VAF and the time-dependent
diffusion coefficient goes actually beyond the polymer translocation problem.
Recall that there are two reasons for the occurrence of an anomalous diffusion
\cite{Bouchaud}: i) The consecutive steps of a random walker are independent but
the waiting time distribution is a sufficiently broad function, $\psi (t) \sim
1/t^{1+\mu}$, where $\mu <1$, so that the first statistical moment {\em does not
exist}. ii) There is a long-time correlation between random steps. In the first
case, the diffusion process could be rationalized in terms of a continuous-time
random walks (CTRW) approach which finally leads to a {\it subdiffusive }
behavior \cite{Sokolov,KM}. The anomalous diffusion in an external force could
be quantified in terms of the {\it fractional} FP-equation
(FFPE)\cite{Sokolov,KM}. Later, this FFPE formalism has been applied to the
driven polymer translocation problem with an external potential approximated by
a function linear in $s$, that is,  $U(s) = U_0 - fs$
\cite{Biophys_Journal,Dubbeldam_3}. Apparently, this approach does not take into
account the tensile force propagation and the moving domain reflecting the chain
reaction which were discussed above (see e.g. the resulting effective nonlinear
potential Eq.~(\ref{U_Potential}) ). The calculations within the FFPE formalism
suggest that to a leading order $\langle s \rangle \propto t^{\mu}$ and $\langle
s^2 \rangle \propto t^{2\mu}$ i.e. the variance goes superdiffusively, $\langle
\Delta s^2 \rangle \propto t^{2\mu}$,  because  $1< 2\mu < 2$
\cite{Dubbeldam_3}. 

Our simulation results make it clear (see Sec.~\ref{Res}  and
Fig.~\ref{fig:dt}a) that in the case of translocation dynamics correlations
persist in time and the CTRW-approach presumably could not be used. In contrast,
the formalism of the V-Langevin equation, used in this Section, gives a general
way to treat the anomalous diffusion. This approach corresponds closely to
driven dynamics because neither the fluctuation-dissipation theorem or the
Stokes-Einstein relation, nor time-translation invariance of the correlation
functions appears to hold in this case.

\section{Simulation Results}
\label{Sim}

\subsection{Model}

In order to verify that the assumptions made in the theoretical analysis are
justified, we performed a number of simulations of polymer chains threading
through a nanopore. We used MD simulations as in Ref.~\cite{Dubbeldam_2}. Here
we briefly recapitulate the used algorithm.

The model we used describes Langevin dynamics of a polymer chain which consists
of $N$ beads that thread through an octagonal pore in a closely-packed wall
(membrane). The interaction between the monomers of the chain is modeled by a
Finitely Extensible Nonlinear Elastic (FENE) springs corresponding to a pair
potential
\begin{align}
 U_{FENE}(r_{ij})&=-\frac{k r_{ij}^2}{2}
\ln\left(1-\frac{r_{ij}^2}{R_0^2}\right),
\end{align}
where $r_{ij}$ is the bond length between two beads and $R_0=1.5$ is the maximal
bond length. All beads  experience excluded volume interactions which are
modeled by the repulsive part of the shifted Lennard-Jones potential, also known
as the Weeks-Chandler-Andersen (WCA) potential. This potential $U_{WCA}$ is
defined by
\begin{align}
 U_{WCA}(r_{ij})=4\epsilon\left[\left(\frac{\sigma}{r_{ij}}\right)^{12}
-\left(\frac{\sigma}{r_{ij}}\right)^{6} + \frac{1}{4}\right] \Theta(r_c - r),
\end{align}
where $\Theta(x)$ is the Heaviside-function, i.e., we use a cut-off
$r_c=2^{-1/6}\sigma$, implying $U_{WCA}=0$ for $r_{ij}>r_c$. The monomers
residing inside the pore experience a constant external force $f$ in the
direction perpendicular to the membrane, which we designate by $x$. The external
force can be implemented by adding a linear potential $U_{ext}$, whose value is
$0$ outside the pore and  $f x$, if $x$ is inside the pore region. Thus, $f$
pulls the chain towards the region of positive $x$ which we refer to as the {\em
trans}-side. The equation of motion for the beads of the chain reads
\begin{align}
 m\frac{d^2\bf{r}_i}{dt^2}=-\nabla \left(U_{FENE}+U_{LJ}+U_{ext}\right) -
\gamma \frac {d {\bf r}_i} {dt} + {\cal R}_i(t), \end{align}
where ${\cal R}_i(t)$ stands for a Brownian random force whose moments obey
$\langle {\cal R}_i(t)\rangle=0$ and $\langle {\cal R}_i^{\alpha}
(t_1) {\cal R}_j^{\beta}(t_2)\rangle = 2 k_B T \gamma
\delta(t_1-t_2)\delta_{\alpha,\beta}\delta_{i,j}$ The parameter values were set
to  $\epsilon=1.0$, $\sigma=1.0$, $k=60.0$, $\gamma=0.70$  and were kept fixed
during the simulations. The temperature had a constant value given by
$T=1.20\epsilon/k_B$.

The membrane is modeled by a plane of beads whose positions are kept fixed.
Eight neighboring monomers are removed to obtain an octagonal pore. The
interaction between the beads of the chain and the plane is mediated through the
repulsive WCA potential. All simulations were performed starting from a
configuration in which initially the chain is placed such that all but one
monomer reside on the {\em cis}-side. In order to prevent the chain from
escaping to the trans-side, we impose reflecting boundary conditions on the
first monomer. The chain is fully translocated when all beads have made their
way to the {\em trans} side.

\subsection{Results}
\label{Res}

\begin{figure}
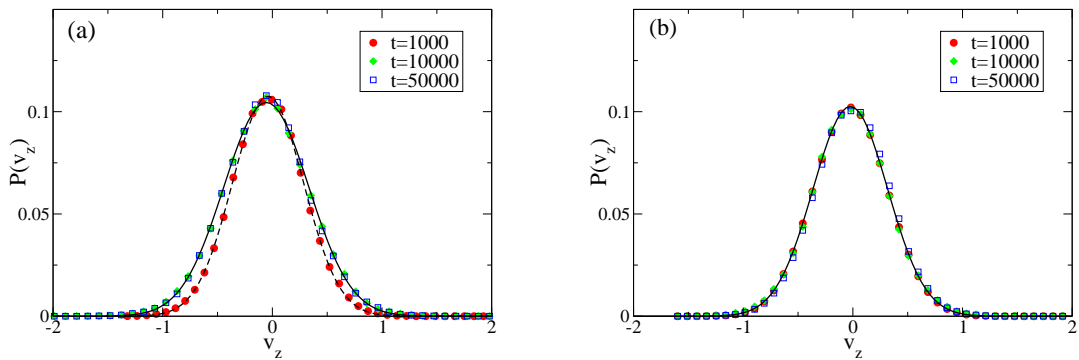

\includegraphics[width=6.5cm]{FIG5a.eps}
\hspace{1cm}
\includegraphics[width=6.5cm]{FIG5b.eps}
\caption{The velocity distribution, $P(v_z)$, for two different forces, (a) $f =
0$, (b) $f = 1$,  and three different time moments. Solid lines represent the
Gaussian distribution fitting.}
\label{Velocity_Distribution}
\end{figure}

First of all we need to prove that the translocation velocity follows indeed a
Gaussian distribution. Simulations were performed for a chain length $N=100$
in the case of free ($f=0$) and driven ($f=1$) translocation. We have
approximated the translocation velocity, $v(t) = d s(t)/dt$, by $v_z (t)$ , the
$z$-component of the Cartesian velocity of the bead inside the pore
\cite{Dubbeldam_2}. Then we recorded data of the velocity at specific time
moments and made histograms over $4000$ runs. As can be seen from Fig.
\ref{Velocity_Distribution}, the velocity corresponds closely to a Gaussian
distribution (solid lines). This justifies our basic assumption, used in the
derivation of the FPE, see Section \ref{Theory}.

\begin{figure}[htbp]
\vspace{0.7cm}

\includegraphics[width=6.0cm]{deltas_N100_F0.eps}
\hspace{1.0cm}
\includegraphics[width=6.0cm]{N100_F1_S2_OLD.eps}
\vspace{0.5cm}

\includegraphics[width=6.0cm]{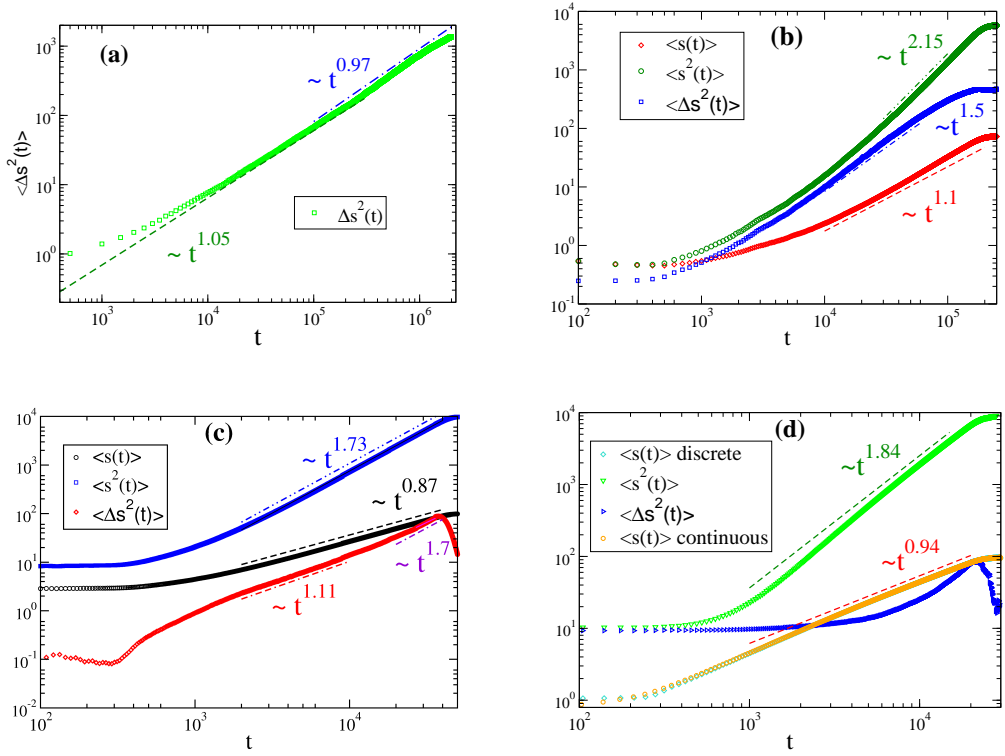}
\hspace{1.0cm}
\includegraphics[width=6.0cm]{s_s2_ds2.eps}
\caption{The first $\langle s \rangle$ and second $\langle s^2 \rangle$
moments as well as the variance $\langle \Delta s^2(t)\rangle = \langle s^2
\rangle - \langle s \rangle^2$ for chain length $N=100$ and different driving
forces $f$: (a)  $f = 0$. (b) $f = 1.0$, (c) $f = 5.0$.  (d) $f = 10.0$.}
\label{fig:figureS2b}
\end{figure}

The statistical moments of the translocation coordinate which are plotted in
Fig. \ref{fig:figureS2b} for different forces, also provide an important
information. While in the unbiased translocation case, Fig.
\ref{fig:figureS2b}a, the diffusion is Brownian (or slightly subdiffusive
\cite{Dubbeldam_2}), in the biased regime the process becomes {\em
superdiffusive}, i.e., the variance $\langle \Delta s^2 \rangle \propto
t^{\theta}$ where $ 1 < \theta < 2$. Moreover, the exponent $\theta$ increases
with the growth of the driving force $f$, namely, $\theta = 1.5$ for $f=1$  (see
Fig. \ref{fig:figureS2b}b), and $\theta = 1.84$ for $f = 10$ (see Fig.
\ref{fig:figureS2b}d). For relatively large forces ($f=5, 10$), the variance
$\langle \Delta s^2 \rangle$ is nonmonotonic, i.e., it goes through a maximum at
a late stage of the translocation. This finding is in agreement with our results
of the FPE numerical solution given in Sec. \ref{Theory_Numerical} (see Fig.
\ref{W_Distributions}). 

\begin{figure}[ht]
 \includegraphics[width =8.0cm]{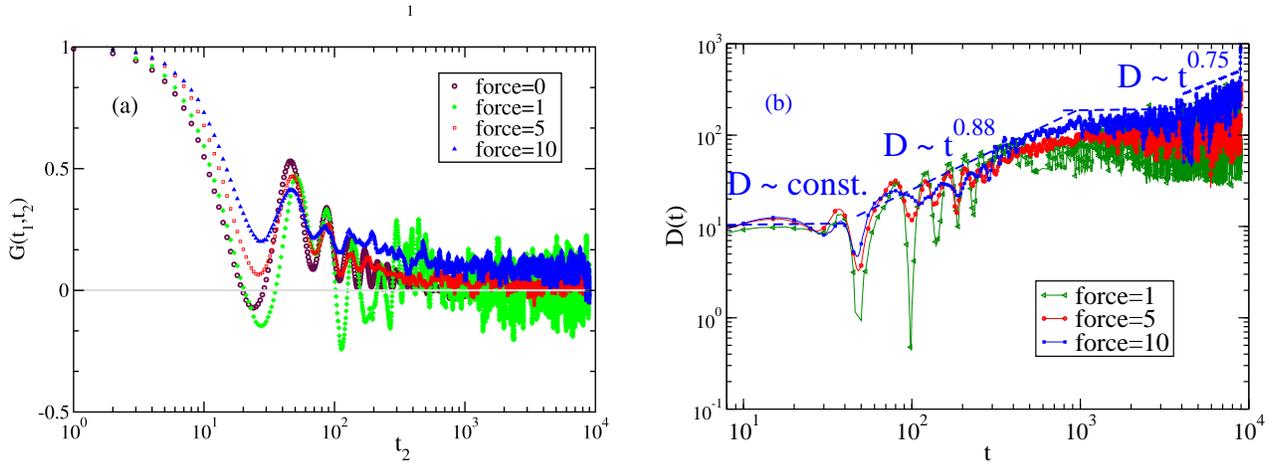}
\hspace{0.5cm}
 \includegraphics[width=8.0cm]{DtF5_10.eps}
\caption{(a) The velocity autocorrelation function for four different values of
force. The stronger forces trigger a longer in time correlations. The vertical 
axis is normalized by $G(t_1, t_1)$.
(b) The diffusion coefficient $D(t)$ as function of time. Larger diffusion
coefficients arise with $f$ increases.}
\label{fig:dt}
\end{figure}

The superdiffusive behavior for $\langle \Delta s^2 \rangle$ implies (according
to the relationship Eq. (\ref{D_2}) ) that the diffusion coefficient is a
growing function of time which has been approximated by a power law, Eq.
(\ref{Running_D}),  in Section \ref{Theory_Numerical}. On the other hand, in
accordance with Eq. (\ref{D}), the time-dependent diffusion coefficient $D(t)$
can be obtained by integration of the VAF $G(t_1, t_2)$. In Fig.~\ref{fig:dt}a
we display the normalized VAF $G(t_1,t_2)/G(t_1,t_1)$ for $t_1=0$ as a function
of $t_2$. This figure indicates that the VAF has a fairly complex behavior: it
changes sign and reveals a long-time tail. It can also be seen that stronger
forces imply longer time correlations which give rise to the observed    
superdiffusive behavior. All measurements have been averaged over $5000$
runs (except for $f=1$ which was obtained by averaging over only $400$ runs).

\begin{figure}
\vspace{1.0cm}

 \includegraphics[width=7.5cm]{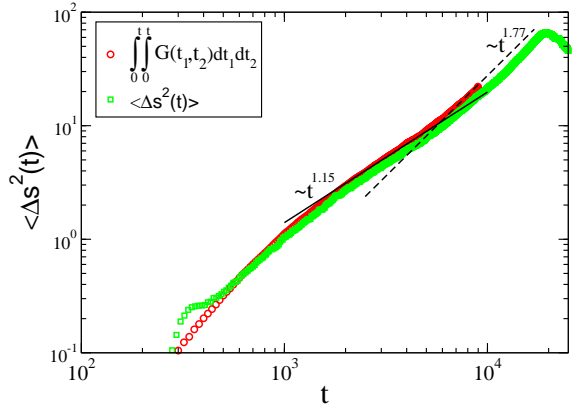}
\caption{The value of $\langle s^2(t) \rangle$ calculated by two different ways:
by integrating the  VAF $G(t_1, t_2)$ over time, i.e. $\langle \Delta s^2
(t)\rangle = \int_{0}^{t} dt_1 \int_{0}^{t} dt_2  G(t_1, t_2)$, and by direct
simulation. Chain length $N = 100$, force $f = 10$.} \label{fig:consistent}
\end{figure}

The corresponding diffusion coefficient $D(t)$ could be obtained by simple
integration over time, according to Eq. (\ref{D}). The result of this
calculation, given in Fig.~\ref{fig:dt}b, shows that the exponent $\gamma$
attains different values within different time intervals. Eventually, we perform
a {\it consistency check}. On the one hand, we integrate the expression for
$G(t_1, t_2)$ over time arguments which provides, according to Eq.
(\ref{Variance}), the value of $\langle \Delta s^2 (t)\rangle = \int_{0}^{t}
dt_1 \int_{0}^{t} dt_2  G(t_1, t_2)$. On the other hand, the MD-simulation
yields the direct time dependence $\langle \Delta s^2 (t)\rangle$. This
comparison of two different calculations of $\langle \Delta s^2 (t)\rangle$,
displayed in Fig. \ref{fig:consistent}, gives almost identical results for
$f=10$.

\section{Conclusion}
\label{Summary}

We have investigated theoretically as well as by extensive MD-simulation the
role of fluctuations for the case of driven polymer translocation through a
nanopore. The consideration is based on the V-Langevin equation,
Eq.~(\ref{Langevin}) where the velocity is a random process with Gaussian
distribution. Keeping this in mind we have derived a corresponding FPE,
Eq.~(\ref{FP_Final}), with a nonlinear drift-term and time-dependent diffusion
coefficient $D(t)$ which can be represented as time integral of VAF $G(t_1,
t_2)$ (see Eq.~(\ref{D})). The derivation requires neither the validity of the
fluctuation-dissipation theorem (as in case of the generalized Langevin equation
approach \cite{Cherail}) nor the time-translation invariance of $G(t_1, t_2)$.
Such a general approach becomes an important tool for the description of driven
translocation dynamics.

It is pertinent to note two limiting cases. In the zero-fluctuation limit, e.g.,
when the driving force is large, one recovers the deterministic solution. For
example, for the first moment it is given by Eq. (\ref{M})) which we have
discussed recently \cite{Dubbeldam_1}. In the zero-force case (nondriven
translocation), in contrast, Eq.(\ref{FP_Final}) leads to the fractional
Brownian motion (fBm) description which was treated in our previous paper
\cite{Dubbeldam_2}. The fact that $D(t)$ is a growing function of time (see Eq.
(\ref{Running_D})), implies that  VAF $G(t_1, t_2)$  has a long-time tail (long
persistence of correlations). This has been verified by our MD-simulation
results (see Fig. \ref{fig:dt}a and Fig. \ref{fig:dt}b ). By making use the
MD-simulation we have also checked the Gaussian distribution of the
translocation velocity (see Fig. \ref{Velocity_Distribution}). 

The numerical solution of the resulting FPE, Eq.~(\ref{FP_Final}), reveals a
number of salient features. If the exponent $\gamma$ governing the time
dependence of the diffusion coefficient, $D(t) \propto t^{\gamma}$ , is large
enough (see the condition given by Eq.~(\ref{Enhancement}) ), fluctuations
``assist`` the translocation dynamics and the translocation coordinate $\langle
s \rangle \propto t^{\beta}$  dependence becomes steeper. Moreover, the variance
follows the law $\langle \Delta s^2 \rangle \propto~t^{\theta}$, where the
exponent $1 < \theta < 2$, i.e., the dynamics is superdiffusive. The increase in
slope due to fluctuations could, therefore, explain the systematic disagreement
between deterministic theory and the MD-simulation results which has been
discussed in ref. \cite{Dubbeldam_1}. One should recall that  the scaling law
for the mean translocation time reads  $\langle \tau \rangle \propto N^{\alpha}$
where $\alpha \approx 1/\beta$. Owing to fluctuations,  $\beta$ grows so that
the translocation exponent $\alpha$ becomes smaller in agreement with the
MD-simulation.

\section*{Acknowledgments}
We thank A.~Y. Grosberg as well as other participants of the CECAM
Workshop ``Polymer Translocation through Nanopores'', held in Mainz on 16-18
September 2012, for fruitful discussions. A. Milchev thanks the Max-Planck
Institute for Polymer Research in Mainz, Germany, for hospitality during his
visit in the institute.  A.~Milchev and V.~G.~Rostiashvili acknowledge support
from Deutsche Forschungsgemeinschaft (DFG), grant No. SFB 625/B4.

\begin{appendix}
\section{Derivation of the Fokker-Planck-Equation (FPE)}
\label{App_1}
Differentiation of Eq. (\ref{PDF}) with respect to $t$ and using the chain rule
and the property of the $\delta$-function yields
 \begin{eqnarray}
 \dfrac{\partial W (s, t)}{\partial t} &=& - \dfrac{\partial}{\partial s}
\left\langle \dfrac{d s(t)}{d t} \delta (s - s(t)) \right\rangle \nonumber\\
&=& - \dfrac{\partial}{\partial s}
\left\langle v \left(s(t) \right) \delta (s - s(t)) \right\rangle
\label{FP_1}
\end{eqnarray}
where the differential operator $\partial / \partial s$ can be put out of
averaging because the $\delta$-function is the only one which depends on $s$.
In Eq. (\ref{FP_1}) we have used the Langevin equation Eq. (\ref{Langevin}).
By making use the relation $\langle v(s(t))\rangle  \delta (s - s(t)) = \langle
v (s)\rangle \delta (s - s(t))$, Eq. (\ref{FP_1}) can be written as
\begin{eqnarray}
\dfrac{\partial W (s, t)}{\partial t}  = - \dfrac{\partial}{\partial s}
 K (s)  \left\langle  \delta (s - s(t)) \right\rangle -
\dfrac{\partial}{\partial s} \left\langle  \delta v (s(t))  \delta (s - s(t))
\right\rangle
\label{FP_2}
\end{eqnarray}
where $\delta v (s(t)) \stackrel{\rm def}{=}  v (s(t)) - \langle v(s(t)) \rangle
$ and we have used Eq. (\ref{First}). The second term in Eq. (\ref{FP_2}) could
be expressed in terms of $W (s, t)$ by employing Novikov's theorem
\cite{Novikov,Zinn-Justin}. According to Novikov's theorem, if $g(t)$ is a
colored Gaussian random process with a zero average, i.e., $\langle g(t)\rangle
= 0$,  and the two-point correlation function is given by $\langle g(t_1)g(t_2) 
\rangle = G(t_1, t_2)$, then for an arbitrary functional,  ${\cal R}
\{g(t)\}$, the average $\left\langle g(t)  {\cal R}\{ g(t)\} \right\rangle  $
can be written as
\begin{eqnarray}
 \left\langle g(t)  {\cal R}\{ g(t)\} \right\rangle  = \int\limits_{0}^{t} \: d
\tau G (t, \tau) \left\langle  \dfrac{\delta  {\cal R}\{ g(t)\} }{\delta
g(\tau)} \right\rangle
\end{eqnarray}
where the symbol $\delta/ \delta g(\tau)$ stands for a functional derivative.

Let's use now the Novikov's theorem  to recast the second term in the r.h.s of
Eq. (\ref{FP_2}). In this case $g(t) = \delta v (s(t))$ and ${\cal R}\{ g(t)\}
= \delta (s - s(t))$. As a result, Eq. (\ref{FP_2}) takes on the form
\begin{eqnarray}
 \dfrac{\partial W (s, t)}{\partial t}  &=& - \dfrac{\partial}{\partial s}
[ K (s)    \: W (s, t)] -  \dfrac{\partial}{\partial s}  \: \int\limits_{0}^{t}
\: d \tau \; G (t, \tau) \left\langle \dfrac{\delta}{\delta v (s(\tau))}
\:  \delta (s - s(t)) \right\rangle \nonumber\\
&=& - \dfrac{\partial}{\partial s} [ K (s)    \: W (s, t)] +
\dfrac{\partial^2}{\partial s^2}  \: \int\limits_{0}^{t} \: d \tau \; G (t,
\tau) \left\langle  \delta (s - s(t)) \dfrac{\delta s (t)}{\delta v (s(\tau))}
\right\rangle
\label{FP_3}
\end{eqnarray}
where the random variable $s(t)$ is treated as a functional of $v (s(\tau))$.
But $s (t) = s (0) + \int_{0}^{t} \: v (s(t')) d t'$ and one has
\begin{eqnarray}
 \dfrac{\delta s (t)}{\delta v (s(\tau))} = \int\limits_{0}^{t} \delta (t' -
\tau) d t' = 1\nonumber
\end{eqnarray}
because $\tau < t$. As a result, one obtains the FPE Eq. (\ref{FP_Final})-
(\ref{D}).

\section{Small noise expansion}
\label{App_2}
In this Appendix, following Section 6.3 of the book of Gardiner \cite{Gardiner},
we give a
short exposition of the small-noise expansion for the FPE,
Eq.(\ref{FP_Final}) where the time-dependent diffusion coefficient $D(t) =
\varepsilon^2 D_0(t)/2$, with $D_0 (t) = (d_0/\tau_0) {\tilde t}^{\gamma}$.
\begin{eqnarray}
\dfrac{\partial W (s, t)}{\partial t}  = - \dfrac{\partial}{\partial s}
[K (s) W (s, t)] + \dfrac{\varepsilon^2}{2} D_0 (t) \:
\dfrac{\partial^2}{\partial s^2}  \: W (s, t)
\label{FP}
\end{eqnarray}
In the same way as in \cite{Gardiner}, one can expand around the
deterministic solution, i.e.,
\begin{eqnarray}
 s = M(t) + \varepsilon x
\end{eqnarray}
where $x$ is a new random variable. Here we discuss only the first
statistical moment which is measured by our computer simulation experiment and
has the following form (to the order $\varepsilon^2$) (see Eq. (6.3.16) in
\cite{Gardiner})
\begin{eqnarray}
  \left\langle s (t) \right\rangle =
M(t) + \varepsilon \left\langle x (t) \right\rangle
 = M (t) + \varepsilon  X_{0}^{1}(t) + \varepsilon^2 X_{1}^{1}(t)
\label{YY}
\end{eqnarray}
where the evolution of $X_{0}^{1}(t)$ and $X_{1}^{1}(t)$ is given by three
ordinary differential equations (see the corresponding Eqs.~(6.3.19), (6.3.20),
(6.3.22) in \cite{Gardiner}) i.e.,
\begin{eqnarray}
\dfrac{d X_{0}^{1}(t)}{d t} &=& {\widetilde K}_{1} (t) \: X_{0}^{1}
(t) \nonumber\\
 \dfrac{d X_{1}^{1}(t)}{d t} &=& {\widetilde K}_{2} (t) \: X_{0}^{2}(t) +
{\widetilde K}_{1} (t) \: X_{1}^{1}(t)\nonumber\\
\dfrac{d X_{0}^{2}(t)}{d t} &=& 2 {\widetilde K}_{1} (t) \: X_{0}^{2}(t) +
D_{0} (t)
\label{XX}
\end{eqnarray}
where ${\widetilde K}_{1} (t)$ and ${\widetilde K}_{2} (t)$ are two expansion
coefficients of $K(s)$ around the deterministic solution, i.e.,
${\widetilde K}_{1}(t) = [d
K (s)/d s]_{s = M(t)}$ and ${\widetilde K}_{1}(t) = (1/2)[d^2 K (s)/d s^2]_{s =
M(t)}$. Taking into account Eq.(\ref{K}), one arrives at
\begin{eqnarray}
 {\widetilde K}_{1}(t) &=& - \dfrac{(1 - \beta)}{\tau_0  \: {\tilde
t}}\nonumber\\
 {\widetilde K}_{2}(t) &=& \dfrac{1}{\tau_0} \left( \dfrac{1-\beta}{2
\beta c_0^\beta} \right) \: \dfrac{1}{{\tilde f}^\beta \: {\tilde t}^{1+\beta}}
\label{KK}
\end{eqnarray}
where the dimensionless force and time are given as ${\tilde f} = a f/T$ and
${\tilde t} = t/\tau_0$, respectively.

In order to solve the first order, linear, inhomogeneous ordinary differential
equations, Eqs. ( \ref{XX}), one recalls that the corresponding generic equation
has the following form
\begin{eqnarray}
 \dfrac{d}{d t} \: y(t) + a (t) \: y (t) = b (t)
\label{Inhomogeneous}
\end{eqnarray}
It easy to verify that this equation has a solution (see \cite{Kamke} Sec.
4.3)
\begin{eqnarray}
 y (t) = \exp\{ - \int a (t) d t\} \left[ \int  \exp\{ \int a (t) d t\}
 \: b (t) dt  + c_2 \right]
 \label{Solution_IH_Equation}
\end{eqnarray}
where $c_2$ is a constant of integration. 

Now let's go back to Eqs.(\ref{XX}).
Solution of the first equation in Eq. (\ref{XX}) is given straightforwardly as
\begin{eqnarray}
 X_{0}^{1} (t) &=& c_1 \; \exp\{ \int_{\tau_0}^t \: {\widetilde K}_{1}(t') dt'\}
= c_1 \; \exp\{ - (1-\beta) \ln (t/\tau_0)\} \nonumber\\
&=& \dfrac{c_1}{{\tilde t}^{1-\beta}}
\label{Solution_of_first}
\end{eqnarray}
where $c_1$ is a constant. Then, taken into account Eq.
(\ref{Solution_IH_Equation}) and  Eq.
(\ref{KK}), the solution for the third equations in Eq.~(\ref{XX}) is given by
\begin{eqnarray}
 X_{0}^{2} (t) &=&  \left(\dfrac{\tau_0}{t} \right)^{2(1-\beta)} \: \left[
\int_{0}^{t} \: \left(\dfrac{t'}{\tau_0} \right)^{2(1-\beta)} \: D_0 (t')
\: d t' \right]\nonumber\\
 &=& d_0 \: {\tilde t}^{1 + \gamma}
\label{X_{0}^{2}}
\end{eqnarray}
It is pertinent to note that while using Eq. (\ref{Solution_IH_Equation}) we
set $c_2 = 0$ because the fluctuation corrections cancel as soon as the
diffusion coefficient is zero, i.e., $X_{0}^{2} (t) = 0$ at $d_0 = 0$.

Again, by making use Eq. (\ref{Solution_IH_Equation}) as well as of Eqs.
(\ref{KK}) and (\ref{X_{0}^{2}}), the solution of the second equation in Eq.
(\ref{XX}) reads
\begin{eqnarray}
 X_{1}^{1} (t) &=& \left(\dfrac{\tau_0}{t} \right)^{(1-\beta)} \: \left[
\int_{0}^{t} \: \left(\dfrac{t'}{\tau_0} \right)^{(1-\beta)} \: {\widetilde
K}_{2}(t') \; X_{0}^{2} (t') \: d t' \right]\nonumber\\
&=& \dfrac{{\widetilde d_0}}{{\widetilde f}^{\beta}} \: \left(\dfrac{\tau_0}{t}
\right)^{(1-\beta)} \: \int_{0}^{t} \: \left(\dfrac{t'}{\tau_0}
\right)^{(1-\beta)} \: \left(\dfrac{\tau_0}{t'} \right)^{(1+\beta)} \:
\left(\dfrac{t'}{\tau_0} \right)^{(1+ \gamma)} \: \dfrac{d
t'}{\tau_0}\nonumber\\
&=& \dfrac{{\widetilde d_0}}{{\widetilde f}^{\beta}} \: {\tilde t}^{1 + \gamma
- \beta}
\label{Solution_of_second}
\end{eqnarray}

Collecting all results, given by Eqs. (\ref{YY}), (\ref{Solution_of_first}) and
(\ref{Solution_of_second}), one has
\begin{eqnarray}
  \left\langle s (t) \right\rangle = \left(c_0
{\tilde f} {\tilde t}\right)^{\beta} + \dfrac{\varepsilon c_1}{{\tilde t}^{1 -
\beta}} + \dfrac{\varepsilon^2 {\tilde d_0}}{{\tilde f}^{\beta}} \: {\tilde
t}^{1 + \gamma - \beta}
\label{First_Moment_Final}
\end{eqnarray}
which goes to Eq. (\ref{YY_Final}) for large times, i.e., ${\tilde t} \gg 1$.

\end{appendix}

\end{document}